\documentclass[sigconf]{acmart}
\AtBeginDocument{%
  \providecommand\BibTeX{{%
    \normalfont B\kern-0.5em{\scshape i\kern-0.25em b}\kern-0.8em\TeX}}}

\usepackage{latexsym}

\usepackage{microtype}

\usepackage{booktabs}
\usepackage{amsmath}
\usepackage{algorithm}
\usepackage{algpseudocode}
\usepackage{multirow}
\usepackage{mathtools}
\usepackage{subfigure}
\usepackage{graphicx}
\usepackage{balance}
\usepackage{color}
\usepackage{enumitem}
\usepackage{xspace}

\renewcommand{\vec}[1]{\ensuremath{\mathbf{#1}}}
\newcommand{\stitle}[1]{\vspace{1mm} \noindent {\bf #1}}
\newcommand{\eg}{{\it e.g.}}

\newcommand{\ie}{{\it i.e.}}
\newcommand{\etc}{{\it etc.}}
\newcommand{\wrt}{w.r.t. }

\newcommand{\bL}{\ensuremath{\mathcal{L}}}

\newcommand{\bP}{\ensuremath{\mathcal{P}}}

\newcommand{\model}{VPGNN}

\copyrightyear{2023}
\acmYear{2023}
\setcopyright{acmlicensed}\acmConference[CIKM '23]{Proceedings of the 32nd
ACM International Conference on Information and Knowledge
Management}{October 21--25, 2023}{Birmingham, United Kingdom}
\acmBooktitle{Proceedings of the 32nd ACM International Conference on
Information and Knowledge Management (CIKM '23), October 21--25, 2023,
Birmingham, United Kingdom}
\acmPrice{15.00}
\acmDOI{10.1145/3583780.3615505}
\acmISBN{979-8-4007-0124-5/23/10}

\begin{document}

\title{Voucher Abuse Detection with Prompt-based Fine-tuning on Graph Neural Networks}

   

\author{Zhihao Wen$^*$}
\email{zhwen.2019@smu.edu.sg}
\affiliation{%
  \institution{Singapore Management University}
   \country{Singapore}
   }

\author{Yuan Fang$^{\dagger}$}
\email{yfang@smu.edu.sg}
\affiliation{%
  \institution{Singapore Management University}
   \country{Singapore}
   }
   
\author{Yihan Liu}
\email{yihan.liu@lazada.com}
\affiliation{%
  \institution{Lazada Inc.}
   \country{Singapore}
   }

\author{Yang Guo}
\email{yg.357086@lazada.com}
\affiliation{%
  \institution{Lazada Inc.}
   \country{Singapore}
   }

\author{Shuji Hao$^{\dagger}$}
\email{hao.shuji@gmail.com}
\affiliation{%
  \institution{Lazada Inc.}
   \country{Singapore}
   }


\thanks{
    $^*$Part of the work done during an internship at Lazada Inc.  
    $^{\dagger}$Corresponding authors.
}

\renewcommand{\shortauthors}{Zhihao Wen, Yuan Fang, Yihan Liu, Yang Guo, \& Shuji Hao}



\begin{abstract}
Voucher abuse detection is an important anomaly detection problem in E-commerce. 
While many GNN-based solutions have emerged, 
the supervised paradigm depends on a large quantity of labeled data.
A popular alternative is to adopt self-supervised pre-training using label-free data, and further fine-tune on a downstream task with limited labels. 
Nevertheless, the ``pre-train, fine-tune'' paradigm is often plagued by the objective gap between pre-training and downstream tasks.
Hence, we propose \model,
a prompt-based fine-tuning framework on GNNs for voucher abuse detection. 
We design a novel graph prompting function to reformulate the downstream task into a similar template as the pretext task in pre-training,
thereby narrowing the objective gap. 
Extensive  experiments on both proprietary and public datasets demonstrate the strength of \model\ in both few-shot and semi-supervised scenarios. Moreover, an online evaluation of \model\ shows a 23.4\% improvement over two existing deployed models.
\end{abstract}



\maketitle

\section{Introduction}
Amidst vigorous market rivalry, user acquisition has been a crucial metric for E-commerce platforms. One of the primary strategies is to introduce electronic \emph{vouchers}, which could help attract more new users or encourage existing users to buy more.  
%
However, the widespread use of vouchers in E-commerce also provides opportunities for abusers. They usually start by registering a massive number of accounts and placing orders just to exploit the vouchers intended for new users. They either resell the goods
(often bought with a deep discount after applying the vouchers) at a higher market price or collude with the sellers by placing orders with vouchers in the sellers' own stores. 
Such behaviors not only cause losses for E-commerce platforms but also damage the ecosystem for legitimate users. 
Thus, it is of great significance to detect orders of these abusive users and prevent them from collecting vouchers. 

\begin{figure}[tbp]
\includegraphics[width=1\linewidth]{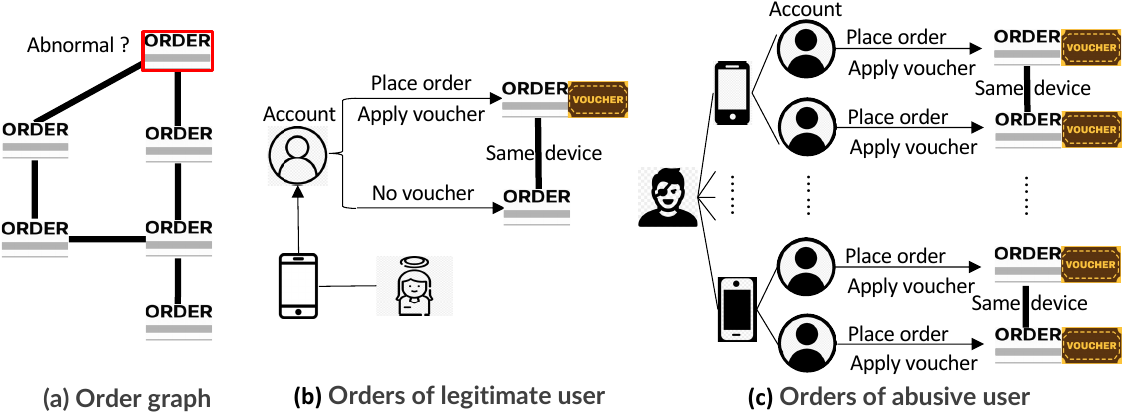}
\vspace{-5mm}
\caption{Illustration of voucher abuse detection.}
\label{fig:background}
\end{figure}

In this paper, we study the problem called \emph{Voucher Abuse Detection}, which aims to detect the orders of abusive users in the E-commerce industry.
A key insight into voucher abuse detection is the network structures among orders. For example, even when orders are placed through different accounts, they can still be related if they share the same device or the same delivery address. 
As illustrated in Fig.~\ref{fig:background}(a), the order graph encodes rich relationships and patterns between orders, which can help differentiate the behaviors of legitimate and abusive users.
As shown in Fig.~\ref{fig:background}(b), a legitimate user typically only logs into one account on one or two 
devices, and applies one voucher 
on their first order to utilize the new buyer incentive. 
In contrast, as shown in Fig.~\ref{fig:background}(c), an abusive user often employs a large number of devices, and in each device, they 
create multiple accounts with multiple sets of information (\eg, e-mail and mobile number). In each account, they collect the voucher for new buyer incentives and just place one order with that voucher. 
By distinguishing the graph structures related to legitimate and abusive orders, we cast the problem of voucher abuse detection as \emph{binary node classification} on graphs.

While voucher abuse detection is a subclass of anomaly detection, existing  solutions  are not ideal choices for the specific problem of voucher abuse detection. On one hand, although traditional machine learning methods \cite{cortes1995support, chen2016xgboost, collobert2004links} are widely used in industry, they do not leverage important graph structure information. Hence, many recent attempts for anomaly detection \cite{ma2021comprehensive} turn to graph neural networks (GNNs) \cite{wu2020comprehensive} to exploit structure information.
On the other hand, voucher abusers often adopt \emph{variable strategies} from time to time 
to reduce detection by the platform. Thus, \emph{timely} and \emph{high-quality} labeled instances of abusive order are crucial for detection, but they are very limited in the production environment of real business scenarios. Hence, most GNN-based approaches cannot cope well in a supervised manner. Meanwhile
self-supervised GNNs \cite{hu2019strategies, hu2020gpt, qiu2020gcc} become promising as they aim to capture intrinsic graph patterns without requiring any annotated label.
However, most self-supervised approaches follow the
``pre-train, fine-tune'' paradigm \cite{hu2019strategies, hu2020gpt},
which suffers from a major drawback: There exists an objective gap between the pre-training and downstream tasks, impairing the generalization of the pre-trained model. 
\stitle{Challenges and our work.}
To bridge the gap between GNN pre-training and downstream tasks, we propose a framework for \underline{V}oucher abuse detection with \underline{P}rompt-based fine-tuning on \underline{G}raph \underline{N}eural \underline{N}etworks (\model). We introduce prompting \cite{liu2023pre,gao2021making} to graph-based tasks, which reformulates the downstream node classification into a similar form as the pretext task in pre-training.
While increasingly popular in NLP, prompt-based learning for GNNs still presents two major challenges. 


First, we cannot directly apply the textual prompting function to bridge various graph-based tasks, 
as textual prompts 
are incompatible with fundamental graph elements (\eg, nodes and edges).
Therefore, in this paper, we propose a \emph{graph prompting function} that reformulates the downstream node classification problem into a pairwise matching task between node tokens and context tokens. Specifically, each context token represents a node class, and node classification is achieved by selecting the most likely class (\ie, context token) that matches the node token.
This pairwise template would be consistent with the paired formulation of many popular pretext tasks in graph pre-training \cite{hamilton2017inductive,hu2020gpt,hu2019strategies, velickovic2019deep}. 


Second, it is still unclear how to initialize the context tokens before any tuning. The context tokens should be 1) informative, to fully exploit prior knowledge learned during pre-training, which is also the initial intention of prompting; 2) robust, especially in low-resource scenarios with limited task labels. 
For informativeness, we reuse the graph readout function from pre-training to initialize the context tokens downstream; for robustness, we augment the limited labeled nodes with their local subgraphs. 
Finally, the context tokens and the pre-trained GNN model are fine-tuned together for downstream classification.

\stitle{Contributions.}
We summarize the contributions of this work.
(1) We propose a novel framework called \model, with a complete pipeline to pre-train and generalize GNNs for voucher abuse detection.
(2) We design a graph prompting function to reformulate the downstream node classification task, which aligns better with the pretext task.
(3) In offline experiments, \model\ can outperform state-of-the-art baselines by up to 4.4\% in the 10-shot setting; in online evaluation, \model\ shows a 23.4\% improvement over two existing deployed models.

\section{Related Work}

Recently, 
GNNs \cite{wu2020comprehensive, kipf2016semi, hamilton2017inductive,velivckovic2018graph, xu2018powerful} have enjoyed widespread application in industry.
Motivated by their success, many GNN-based anomaly detection algorithms have emerged, including GAS \cite{li2019spam}, FdGars \cite{wang2019fdgars}, GraphConsis \cite{liu2020alleviating} and CARE-GNN \cite{dou2020enhancing} for review fraud detection, GeniePath \cite{liu2019geniepath} and SemiGNN \cite{wang2019semi} for financial fraud detection,
FANG \cite{nguyen2020fang} for fake news detection, ASA \cite{wen2020asa} for mobile fraud detection, and MTAD-GAT \cite{zhao2020multivariate} for time-series anomaly detection.
There are also some unsupervised anomaly detection GNNs \cite{jin2021anemone, liu2021anomaly, ding2019deep, bandyopadhyay2019outlier}, but they are often less reliable since they do not make use of any labeled data.
To reduce labeling requirements, 
GNN pre-training \cite{hu2019strategies, qiu2020gcc, hu2020gpt, lu2021learning, liu2022graph} has become a popular alternative, which aims to capture general patterns on label-free graphs. 
Nevertheless, a considerable amount of labeled data are still required for fine-tuning on downstream tasks.

In NLP, prompting \cite{liu2023pre} has emerged to overcome the objective gap between the pretext  and downstream tasks.
Prompting reformulates the downstream task to follow a similar template as the pretext task so that the downstream task can be optimized with a light tuning or even without tuning \cite{brown2020language}.
On graphs, there have also been some attempts to leverage prompt-based learning. GPF \cite{fang2022prompt} only trains a prefix prompt vector appended to the node features, lacking a unified template for pretext and downstream tasks.
GPPT \cite{sun2022gppt} and GraphPrompt \cite{liu2023graphprompt} attempt to unify the pretext task of link prediction and downstream classification, but the unification is incompatible with voucher abuse detection, where abusive orders are linked with many legitimate orders. 

\section{Preliminaries} 


\stitle{Pre-training.} 
Many pretext tasks \cite{hamilton2017inductive, hu2019strategies, velickovic2019deep} have been proposed to pre-train GNNs.
In voucher abuse detection, abusive orders are the minority and the majority are legitimate orders. Hence, we utilize DGI \cite{velickovic2019deep} to maximize the local-global mutual information, whereby the graph-level global information captures the ``normal'' patterns manifested by the majority, which helps indicate the extent of a node deviating from what is normal.

Specifically, let $\vec{H}$ be the node representation matrix in which each row $\vec{h}_i$ is the representation of node $i$ generated by a GNN encoder, parameterized by $\theta$. Furthermore, let $\vec{h}_G$ be the global representation of $G$, given by $\vec{h}_G=\Omega(\vec{H}; \omega)=\operatorname{POOL}(\{\vec{h}_i \mid i \in V\})$ where $\Omega$ is a readout function with parameters $\omega$.
In pre-training, DGI \cite{velickovic2019deep} aims to minimize the following loss:
\begin{align}
    \textstyle\arg\min_{\theta, \omega, \phi} \textstyle\sum_{(i, G)} \mathcal{L}^{\text{pre}}\left(\Phi^{\text{pre}}\left(\vec{h}_i, \vec{h}_G;\phi\right), \operatorname{match}(i, G)\right),
    \label{eq:l_pre}
\end{align}
where $\Phi^{\text{pre}}$, parameterized by $\phi$, is a projection head to evaluate the matching score of a node-graph pair $(i,G)$, which measures the local-global consistency. 
Note that the supervision comes from
$\operatorname{match}(i, G)$, which is an indicator function: 1 if node $i$ is from the original graph $G$;
0 if node $i$ is from a corrupted version of $G$. 
Hence, the pre-training process does not require any human annotation, updating model parameters including $\theta$ (GNN), $\omega$ (readout), and $\phi$ (projection head) in a self-supervised manner.

\stitle{Fine-tuning.} 
The pre-trained GNN parameters $\theta^\text{pre}$, optimized by Eq.~\eqref{eq:l_pre}, serve as a good initialization for downstream classification tasks. Following the ``pre-train, fine-tune'' paradigm, the initialization is further fine-tuned together with
a new set of classification weights $\psi$, by optimizing the following:
\begin{align}
   \textstyle \arg\min_{\theta', \psi} \textstyle \sum_{i \in V} \mathcal{L}^{\text{down}}\big(\Phi^{\text{down}}(\vec{h}_i;\psi), y_i\big),
    \label{eq:l_down}
\end{align}
where $\Phi^\text{down}$ is a new projection head with randomly initialized parameters $\psi$, replacing the pretext projection head $\Phi^\text{pre}$. $\mathcal{L}^{\text{down}}$ is the downstream classification loss (\eg, cross entropy), and  $y_i$ is the task-specific label of node $i$. 

\begin{figure}[t]
   \includegraphics[width=1\linewidth]{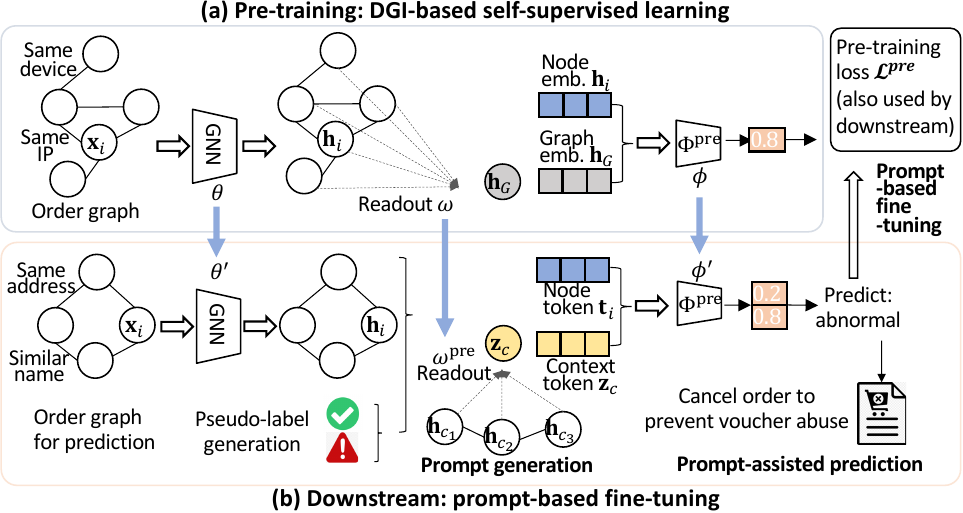}
  \vspace{-5mm}
  \caption{Overall framework of \model.}
  \label{fig:frame}
\end{figure}
%

\section{Proposed Approach}
In this section, we give an overview of our approach and elaborate on our prompt-based fine-tuning. 

\subsection{Overview of \model}
Unlike textual prompts in NLP,
on graph data, it is non-trivial to design the prompts due to the incompatibility between traditional textual prompts and graph elements. 
To materialize prompt-based fine-tuning on graphs, our \model\ has two major stages, as shown in Fig.~\ref{fig:frame}. First, in Fig.~\ref{fig:frame}(a), we conduct pre-training based on DGI. Then, in (b), we perform prompt-based fine-tuning for the downstream voucher abuse detection. 

In particular, prompt-based fine-tuning consists of three key modules: (1) \emph{Prompt generation}: Our graph prompting function generates a set of token pairs for each input node; (2) \emph{Prompt-assisted prediction}: For each input node, the matching probabilities of its token pairs can be scored by the same projection head in pre-training. The matching probabilities will be used for the prediction of legitimate/abusive orders. (3) \emph{Prompt-based fine-tuning}: The context tokens in the prompt will be fine-tuned together with the pre-trained GNN, and the key difference from traditional fine-tuning is that the same pretext projection  head $\Phi^\text{pre}$ and pretext loss function $\bL^\text{pre}$ employed in pre-training  will be \emph{reused} for the downstream task, without needing a new projection head or task loss. 

In the following, we will first briefly introduce the data preparation process, then focus on  the downstream stage in Fig.~\ref{fig:frame}(b), as the pre-training stage has been introduced in the preliminaries. 

\subsection{Data preparation} \label{sec:method:data}
Our model \model\ requires two key input elements, including an order graph and a small number of labels, as we lay out below.

\stitle{Graph construction.} We work with two categories of proprietary raw data: (1) User profiles with details such as email, IP and shipping addresses, and (2) Buyer journey logs such as logins, orders, payments, \etc\ 
We construct an order graph based on various shared attributes like the same device or address, similar usernames, \etc, as shown in Fig.~\ref{fig:background}(a), to capture collusive patterns between orders.
Note that different order graphs could be created based on different markets or days, where pre-training may be conducted on one order graph for downstream prediction on a different order graph, enabling generalization across markets or time.

\stitle{Limited pseudo-label generation.} Due to the fluid nature of voucher abuse, timely and high-confidence labels are preferred. 
Hence, we generate pseudo-labels by employing a set of predefined
business rules, which have been crafted by Lazada Inc.’s internal experts.  
In particular, the rules are designed to be conservative to avoid upsetting legitimate users, so that only the most conspicuous abusive orders would be flagged out.  
Hence, these pseudo-labels hold high confidence, but their availability is limited. These few labeled examples will be employed for our prompt-based fine-tuning in the downstream prediction.



\subsection{Prompt generation, prediction, and tuning}

Instead of introducing a new projection head and loss for downstream task, our prompt-based fine-tuning framework generates and tunes prompts, and makes predictions based on the prompts.

First, we propose a graph prompting function $\bP$, which transforms an input node $i$ into a prompt $\vec{p}_i$. The prompt $\vec{p}_i$ is a continuous embedding vector and has the same shape as the input to the pretext projection head in pre-training. Given that our pretext task in DGI is to maximize the mutual information in node-graph pairs, our prompt $\vec{p}_i$ also assumes a pairwise template, consisting of a pair of node token and context token. For a node $i\in V$, we have 
\begin{align}
    \vec{p}_i=\bP(i)=[\vec{t}_i, \vec{z}_c],
    \label{eq:prompt}
\end{align}
where the node token $\vec{t}_i$ is a vector representation of  node $i$ that can be encoded by a GNN, and the context token $\vec{z}_c$ is a learnable embedding for class $y_c$ in the downstream task, as explained below.  

\stitle{Node token.}
The node token captures the node information. Corresponding to the input text embedding in NLP, our node token $\vec{t}_i$ is an embedding of the input node $i$. It can simply be $\vec{h}_i$, the representation of node $i$ as encoded by a GNN,
or an aggregation of the embedding vectors of adjacent nodes of $i$. In our work, we simply implement $\vec{t}_i = \vec{h}_i$, which is the direct output of the GNN.

\stitle{Context token.}
The context token is designed to capture the contextual information about a class.
Inspired by prompt tuning \cite{li2021prefix, liu2022p}, we model our context token with a learnable vector $\vec{z}_c$ for each class $y_c$ in the downstream task. Suppose there is a set of $C$ classes $\{1,2, \dots, C\}$.
Hence, for each input node $i$, we can pair its node token with $C$ different context tokens, to form $C$ token pairs, namely, $(\vec{t}_i, \vec{z}_1), (\vec{t}_i, \vec{z}_2), \ldots, (\vec{t}_i, \vec{z}_{C})$.
The context tokens can be represented by a learnable prompt matrix
$\vec{Z} = [\vec{z}_1, \vec{z}_2, \ldots, \vec{z}_{C}]^\top \in \mathbb{R}^{C\times d}$,
where $d$ is also the embedding dimension of node representations. 


\stitle{Prompt initialization.}
Since the context tokens are learnable vectors, we need to address their initialization. 
The traditional way is random initialization, \ie, trained from scratch, which is uninformative and fails to exploit our pre-trained model.

An optimal context token $\vec{z}_c$ is designed to capture the contextual information about class $c \in \{1,\ldots,C\}$. One straightforward option is to take the mean of the representation vectors of all labeled nodes in class $c$, since information about the class should ideally relate to class centroid. This is more informative than random initialization, by making use of the embedding vectors encoded by the pre-trained GNN. However, it still suffers from two problems. Firstly, in low-resource scenarios where each class has very few labeled nodes, the sample mean could be an unreliable estimation of the class centroid. Secondly, a simple mean cannot capture complex non-linear relationship between the contextual information about the class and the class centroid.   

To improve the robustness of the initialization, we augment the labeled nodes with their neighboring nodes, which can provide additional local information around a labeled node. To improve the informativeness, we reuse the graph Readout function from pre-training. As the Readout is designed to pool the nodes in a graph to derive a graph-level summary representation, it can be similarly applied downstream to pool the nodes related to a class to derive a class-level summary representation. The class-level summary can serve as a more informative initialization for the context token, which aims to capture information about the class.
Specifically, for each class $c$,
we gather the set of labeled nodes and their neighboring nodes $V_c=\{c_1,c_2,\ldots,c_N\}$, where $N=|V_c|$. Then, we construct an input embedding matrix $\vec{H}_c=[\vec{h}_{c_1}, \vec{h}_{c_2}, \ldots, \vec{h}_{c_N}]^\top \in\mathbb{R}^{N\times d}$ for the nodes in $V_c$, where $\vec{h}_{c_i}$ is the representation of node $c_i \in V_c$ as encoded by the pre-trained GNN. Subsequently, the context token $\vec{z}_c$ can be initialized by $\Omega(\vec{H}_c; \omega^\text{pre})$,
where $\omega^\text{pre}$ is the pre-trained parameters of the Readout function. For efficiency, we only sample $\eta$ neighbors from each labeled node.

Based on our prompt design, we outline prompt-based learning for voucher abuse detection, which involves prompt-assisted prediction and prompt-based fine-tuning.

\stitle{Prompt-assisted prediction.}
For each input node $i$, the prompting function generates two token pairs
$(\vec{t}_i, \vec{z}_0)$ and $(\vec{t}_i, \vec{z}_1)$, given classes $\{0=\text{legitimate},1=\text{abusive}\}$ for the binary voucher abuse detection.
We can leverage the same pretext projection head $\Phi^\text{pre}$ to score the matching probability of each token pair, similar to scoring the node-graph pairs in pre-training. Thus, we predict the order represented by node $i$ as abusive if and only if 
\vspace{-1mm}
\begin{align}
\Phi^\text{pre}(\vec{t}_i, \vec{z}_1;\phi') > 
\Phi^\text{pre}(\vec{t}_i, \vec{z}_0;\phi'),
\label{eq:predict}
\end{align}
since $\vec{z}_0,\vec{z}_1$ capture the contextual information about the two classes. Here $\phi'$ can be initialized by $\phi^\text{pre}$, the pre-trained parameters of the projection head, and further fine-tuned based on task-specific labels as we show next.

\stitle{Prompt-based fine-tuning.}
Our prompt design allows us to reuse not only  the pretext projection head without introducing a new classification head,  but also the pretext task loss without formulating a new task loss.
\begin{align}
  \textstyle  \arg\min_{\theta', \phi', \vec{Z}} \textstyle\sum_{(i, c)} \mathcal{L}^{\text{pre}}(\Phi^\text{pre}(\vec{t}_i, \vec{z}_c;\phi'); \text{match}(i, c))
    +\lambda \mathcal{L}^o,
    \label{eq:overall}
\end{align}
where $\text{match}(i, c)$ is a reloaded indicator function: 1 if node $i$ is labeled $c$; $0$ otherwise.  $\mathcal{L}^o=\left\|\vec{Z}\vec{Z}^\top-\vec{I}\right\|_2^2$ is the orthogonal constraint on the prompt matrix to promote separability of each class, and $\lambda\ge 0$ is a co-efficient to control the importance of the constraint.

Compared to the traditional fine-tuning in Eq.~\eqref{eq:l_down}, we still optimize \wrt~the pre-training loss function $\mathcal{L}^{\text{pre}}$ with the same projection head $\Phi^\text{pre}$. 
Meanwhile, we optimize the prompt matrix $\vec{Z}$ in Eq.~\eqref{eq:overall} instead of the new classification weights $\psi$ in Eq.~\eqref{eq:l_down}. Hence, our prompt-based approach unifies the pretext and downstream task, narrowing the gap between pre-training and downstream objectives,
where $\theta^{\prime}$ and $\phi^{\prime}$ are the parameters of pre-trained GNNs and projection head, $\vec{Z}$ are the well-initialized context tokens, $\lambda$ is the hyper-parameter, controlling the orthogonal constraint.

\section{Experiments}\label{sec:offline}
We conduct a comprehensive suite of offline experiments first, followed by an online evaluation of our deployed model. 

\subsection{Offline setup and results}

\stitle{Datasets.}
First, we collect four proprietary large-scale datasets for voucher abuse detection, named
\textbf{VN0909}, \textbf{VN1010}, \textbf{ID0909}, and \textbf{ID1023}, from an e-commerce platform provided by Lazada Inc.~on the premise of complying with all the security and privacy policies. Each dataset is a graph with millions of nodes, where the nodes represent the orders with pre-defined features, and the edges are pre-defined relationships between them. VN0909 and VN1010 are from Lazada's Vietnam market, collected on Sep.~9 and Oct.~10 2022, while ID0909 and ID1023 are from Lazada's Indonesia market, collected on Sep.~9 and Oct.~23. Note that VN0909 is only used for \emph{pre-training}, and we do test on three other  datasets. 
Second, 
we also use a public dataset on anomaly detection, namely \textbf{Amazon} \cite{dou2020enhancing}. We perform pre-training on itself.


\begin{table*}[t]
\caption{Performance comparison between \model\ and the baselines, in percent, with 95\% confidence intervals.} \label{table:main results}
{
\footnotesize 
\vspace{-2mm}
In each row, the best result is bolded and the runner-up is underlined. 
``$/$'' indicates no result obtained  due to out-of-memory issue or excessively long training time (>72 hours).\vspace{1mm}}
\footnotesize
\centering
\begin{tabular}{@{}c|ccc|ccc|ccc|ccc|c@{}}
\hline
\multicolumn{1}{c|}{} & SVM & XGBoost & MLP & GCN & SAGE$_\text{sup}$  &GAT& CARE-GNN &GeniePath & AMNet & DCI &SAGE$_\text{unsup}$  & Pre-train & \model\  \\ \hline

\multicolumn{14}{c}{\textit{\textbf{Number of shots = 10}}} \\ \hline
\multirow{1}{*}{VN1010} & 37.1\textpm{8.9}  & \underline{65.7}\textpm{5.5} &62.9\textpm{2.8} 
&59.1\textpm{6.7}&61.9\textpm{3.6} & 60.9\textpm{4.6}
& /\ &58.0\textpm{4.5}&/\ &/\ & 61.8\textpm{4.6}  &64.8\textpm{3.6} 
&\textbf{67.1}{}\textpm{3.1}  
\\ 

\multirow{1}{*}{ID0909} & 28.1\textpm{11.0}&51.3\textpm{9.9} &61.6\textpm{3.8} 
&64.1\textpm{3.9}&61.2\textpm{7.3}  & 65.6\textpm{3.7}
&/\ &62.1\textpm{2.8}&/\  &/\ & 62.2\textpm{3.7} & \underline{66.1}\textpm{3.0}  
&\textbf{69.0}{}\textpm{3.7}  
\\ 

\multirow{1}{*}{ID1023} & 38.7\textpm{8.3} &73.5\textpm{6.1} &69.3\textpm{2.1} 
&69.3\textpm{4.8}&71.3\textpm{5.2} & \underline{73.7}\textpm{3.6}
&73.0\textpm{2.9}&72.0\textpm{5.0}&70.0\textpm{3.5} &73.4\textpm{1.6} & 67.5\textpm{5.3} &71.8\textpm{5.2} 
&\textbf{75.1}{}\textpm{1.9} 
\\ 

\multirow{1}{*}{Amazon} 
& 41.4$\pm$9.2 &  62.5$\pm$11.5 &63.3 $\pm$5.7
& 16.5$\pm$4.9 & 59.9$\pm$9.1 &20.5$\pm$6.1
& 38.6$\pm$2.9 & 30.7$\pm$2.8 & \underline{64.8}$\pm$6.2
& 18.5$\pm$4.0 & 36.7$\pm$6.0 & 62.3$\pm$8.1
& \textbf{70.0}{}$\pm$2.7
\\ \hline

\multicolumn{14}{c}{\textit{\textbf{Number of shots = 20}}} \\ \hline
\multirow{1}{*}{VN1010} & 59.2\textpm{3.8} &73.1\textpm{3.9} &69.2\textpm{2.2} 
&71.6\textpm{2.8}&73.6\textpm{3.2}  & 74.5\textpm{4.4}
&/\ &69.5\textpm{5.7}&/\ 
&/\ & 72.8\textpm{2.0} &\underline{75.7}\textpm{3.0}  
&\textbf{75.9}\textpm{2.8}  
\\ 

\multirow{1}{*}{ID0909} & 53.1\textpm{6.1} &64.9\textpm{3.8} &64.9\textpm{1.7} 
&68.7\textpm{2.5}&70.3\textpm{2.8}  & 71.0\textpm{3.4}
&/\ &61.4\textpm{9.1}&/\ 
&/\ & 67.5\textpm{2.2} &\underline{71.4}\textpm{2.5}  
&\textbf{72.7}{}\textpm{2.8}  
\\ 

\multirow{1}{*}{ID1023} & 65.2\textpm{3.6} &78.9\textpm{1.4} &74.7\textpm{1.6} 
&79.0\textpm{1.7}&\underline{81.5}\textpm{1.3} & 81.1\textpm{1.1}
&74.2\textpm{0.9}&80.7\textpm{4.2}&75.6\textpm{1.9} 
&79.4\textpm{1.3} & 78.1\textpm{2.1} &81.3\textpm{1.4} 
&\textbf{81.8}\textpm{1.1}  
\\ 

\multirow{1}{*}{Amazon} 
& 60.3$\pm$3.6 & 72.9$\pm$8.2 & 70.4$\pm$4.4
& 16.8$\pm$8.0 & 63.0$\pm$8.5 & 48.8$\pm$10.1
& 42.2$\pm$6.7 & 30.8$\pm$4.4 & \underline{75.1}$\pm$3.1
& 21.8$\pm$2.6 & 54.5$\pm$2.8 & 73.1$\pm$3.9
& \textbf{76.6}$\pm$2.5 
\\ \hline

\multicolumn{14}{c}{\textit{\textbf{Semi-supervised}}} \\ \hline
\multirow{1}{*}{VN1010} & 86.7\textpm{0.1}  &87.8\textpm{0.1} &86.7\textpm{0.1} 
&91.8\textpm{0.1}&\underline{94.1}\textpm{0.1} & 91.9\textpm{0.0}
&/\ &91.7\textpm{0.4}&/\  
&/\ & 89.3\textpm{0.0} &\underline{94.1}\textpm{0.1}   
&\textbf{95.2}{}\textpm{0.1}  
\\ 

\multirow{1}{*}{ID0909} & 86.0\textpm{0.2} &89.2\textpm{0.3} &86.8\textpm{0.3} 
&92.2\textpm{0.2}&\underline{93.4}\textpm{0.2} & 92.3\textpm{0.2}
&/\ &91.1\textpm{0.4}&/\ 
&/\ & 86.3\textpm{0.2} &93.3\textpm{0.2}  
&\textbf{94.1}{}\textpm{0.2}  
\\ 

\multirow{1}{*}{ID1023} & 89.3\textpm{0.1} &89.9\textpm{0.2} &88.8\textpm{0.2} 
&94.4\textpm{0.1}&\underline{95.6}\textpm{0.1}  & 94.5\textpm{0.1}
&87.0\textpm{0.3}&94.5\textpm{0.1}&94.1\textpm{0.3} 
&93.7\textpm{1.0} & 92.6\textpm{0.1} &\underline{95.6}\textpm{0.1} 
&\textbf{96.2}{}\textpm{0.1}  
\\ 

\multirow{1}{*}{Amazon} 
& 78.8$\pm$1.1 & 75.6$\pm$2.7 & 78.1$\pm$1.8
& 34.4$\pm$3.6 & 81.1$\pm$1.5 & 73.3$\pm$2.9
& 45.2$\pm$5.6 & 30.9$\pm$4.5 & \textbf{81.7}$\pm$1.1
& 26.1$\pm$4.5 & 76.1$\pm$0.9 & \underline{80.9}$\pm$0.9
& 80.6$\pm$0.7
\\ \hline
\end{tabular}
\end{table*}

\stitle{Task setup.}
We construct downstream tasks in \emph{10-shot}, \emph{20-shot} and semi-supervised settings. 
In 10- or 20-shot settings, we sample 10 or 20 nodes labeled as anomaly for training, respectively.
In the semi-supervised setting, we sample 100 anomaly labels on Amazon and 5000 anomaly labels on others for training.
As anomaly detection is an imbalanced binary classification problem, we further sample negative nodes (\ie, normal nodes without anomaly) for training so that the anomaly rate in training is similar to the overall rate.
For all  settings, a validation set with an equal size to the training set is also sampled, and the remaining will be used for testing. Finally, we randomly generate 10 different splits of train/validation/test in each setting,
and report the average performance with a 95\% confidence interval. 

\stitle{Baselines.}
We consider competitive baselines from four categories. 
(1) \emph{Classical machine learning}, widely used in industry,
including \textbf{SVM} \cite{cortes1995support}, \textbf{XGBoost} \cite{chen2016xgboost}, and \textbf{MLP} \cite{collobert2004links};
(2) \emph{General semi-supervised GNNs}, which use both node features and graph structures and train end to end, including \textbf{GCN} \cite{kipf2016semi}, \textbf{GAT} \cite{velivckovic2018graph}, and \textbf{SAGE$_\text{sup}$}, the supervised version of GraphSAGE \cite{hamilton2017inductive};
(3) \emph{Anomaly detection GNNs}, which are special-purpose GNNs designed for anomaly detection and trained in a supervised fashion.
\textbf{CARE-GNN} \cite{dou2020enhancing} enhances the GNN aggregation process with three unique modules against camouflages.
\textbf{GeniePath} \cite{liu2019geniepath} is  an approach learning adaptive receptive fields of GNN, with an adaptive path layer consisting of two complementary functions designed for breadth and depth exploration respectively.
\textbf{AMNet} \cite{ijcai2022p270} is an adaptive multi-frequency GNN, capturing both low-frequency and high-frequency signals, and adaptively combine signals of different frequencies;
(4) \emph{Pre-trained GNNs}, which perform pre-training on label-free graphs, including
\textbf{DCI} \cite{wang2021decoupling}, \textbf{SAGE$_\text{unsup}$}, the unsupervised version of GraphSAGE \cite{hamilton2017inductive} which employs a form of linear probe that is known to be a strong few-shot learner \cite{tian2020rethinking}, and
\textbf{Pre-train} which uses the same pre-training strategy as  \model, and is then fine-tuned with a new projection head.  


\stitle{Settings.}
We set the number of hidden units to 128, using two layers with  ReLU activation for all GNNs, except for GeniePath and CareGNN. Geniepath uses 16 hidden units and 7 layers, while CareGNN has 64 hidden units and 1 layer, as recommended in  \cite{liu2019geniepath, dou2020enhancing}. 
The Adam optimizer is applied to both pre-training and fine-tuning. The learning rate is 0.01 on the Lazada datasets and 0.001 on Amazon. For \model, we use a 128$\times$128 fully connected layer as the Readout function, inner product as the pretext task projection head, set $\eta=5$ for the number of neighbors sampled per labeled node for context token initialization, and $\lambda=0.01$ for the coefficient of the orthogonal constraint.


\stitle{Classification performance.}
In Tab.~\ref{table:main results}, we compare the performance of \model\ with the baselines.
First, given that pre-training is done on VN0909, the superior performance of \model\ on VN1010, ID0909 and ID1023 shows its generalization ability across time and/or markets. Furthermore, \model\ is a strong few-shot learner, as it attains larger improvements relative to the runner-up under the 10-shot setting. In general, other pre-trained GNNs also tend to perform better under the few-shot settings.
Second, classical  machine learning methods are generally inferior to GNN-based methods, demonstrating that graph structures can complement node features in our problem. 
Third, those GNNs specifically designed for anomaly detection only achieve similar performance to vanilla GNNs on the Lazada datasets, showing that these special-purpose GNNs cannot play to their strengths in voucher abuse detection due to the fluidity and complexity of the problem.

\begin{figure}[t] 
   \begin{minipage}[t]{0.52\linewidth}
    \centering
    \footnotesize
    \addtolength{\tabcolsep}{-1pt}
    \renewcommand*{\arraystretch}{1.6} 
    \vspace{-24mm}
    \captionof{table}{Ablation study on\\ID1023 under different shots.}
    \vspace{-3mm}%
    \hspace{-3mm}%
        \begin{tabular}{c|c|c|c}  
    \hline
     Model \textbackslash\ shots&10&20&Semi \\\hline
    No prompt &71.82&81.31&95.61 \\
      Random init.&73.76&79.07&95.55\\
            No constr.&74.72&81.81&96.12\\
    \hline
    \model &\textbf{75.09}&\textbf{81.84}&\textbf{96.21}\\
    \hline
\end{tabular}
    \label{tab:abla_shots} 
  \end{minipage}%
  \begin{minipage}[t]{0.47\linewidth} 
      \centering 
      \setlength{\abovecaptionskip}{0.28cm}
      \includegraphics[width=\textwidth]{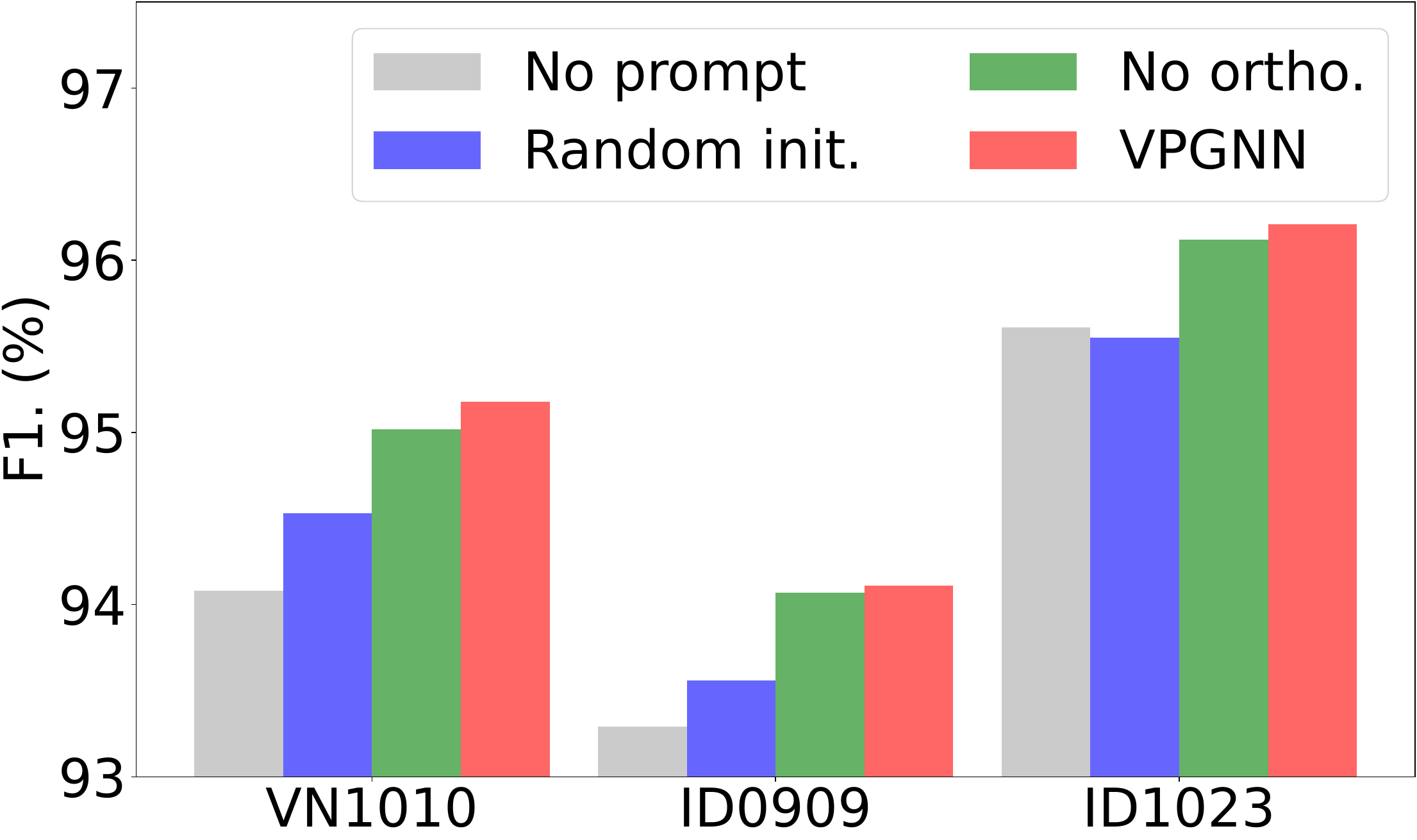}
      \vspace{-5mm}
      \captionof{figure}{Ablation study on semi-supervised setting.}
      \label{fig:abla_dataset}
   \end{minipage}%
\end{figure}

\stitle{Ablation study.}
To better understand the contribution of each component in \model, we compare \model\ with the following ablated models. 1) \emph{No prompt}, which follows the traditional ``pre-train, fine-tune'' paradigm without prompting; 2) \emph{Random init.}, which randomly initializes the context tokens; 3) \emph{No constr.}, which removes the orthogonal prompt constraint in Eq.~\eqref{eq:overall}. We report the results on ID1023 with different shots in Tab.~\ref{tab:abla_shots}, and under the semi-supervised setting on different datasets in Fig.~\ref{fig:abla_dataset}. 
%
 \model\ 
outperforms all the ablated models consistently, demonstrating the
overall benefit of integrating various components. Among the ablated models,
\emph{No prompt} performs rather poorly in most cases, especially under the 10-shot setting. This demonstrates the superiority of prompt-based fine-tuning compared with traditional fine-tuning in low-resource setting. \emph{Random init.} also performs not as well, showing that the token initialization in \model\ is crucial.  



\begin{figure}[t]
  \begin{minipage}[t]{0.48\linewidth}
    \centering
    \includegraphics[width=1\linewidth]{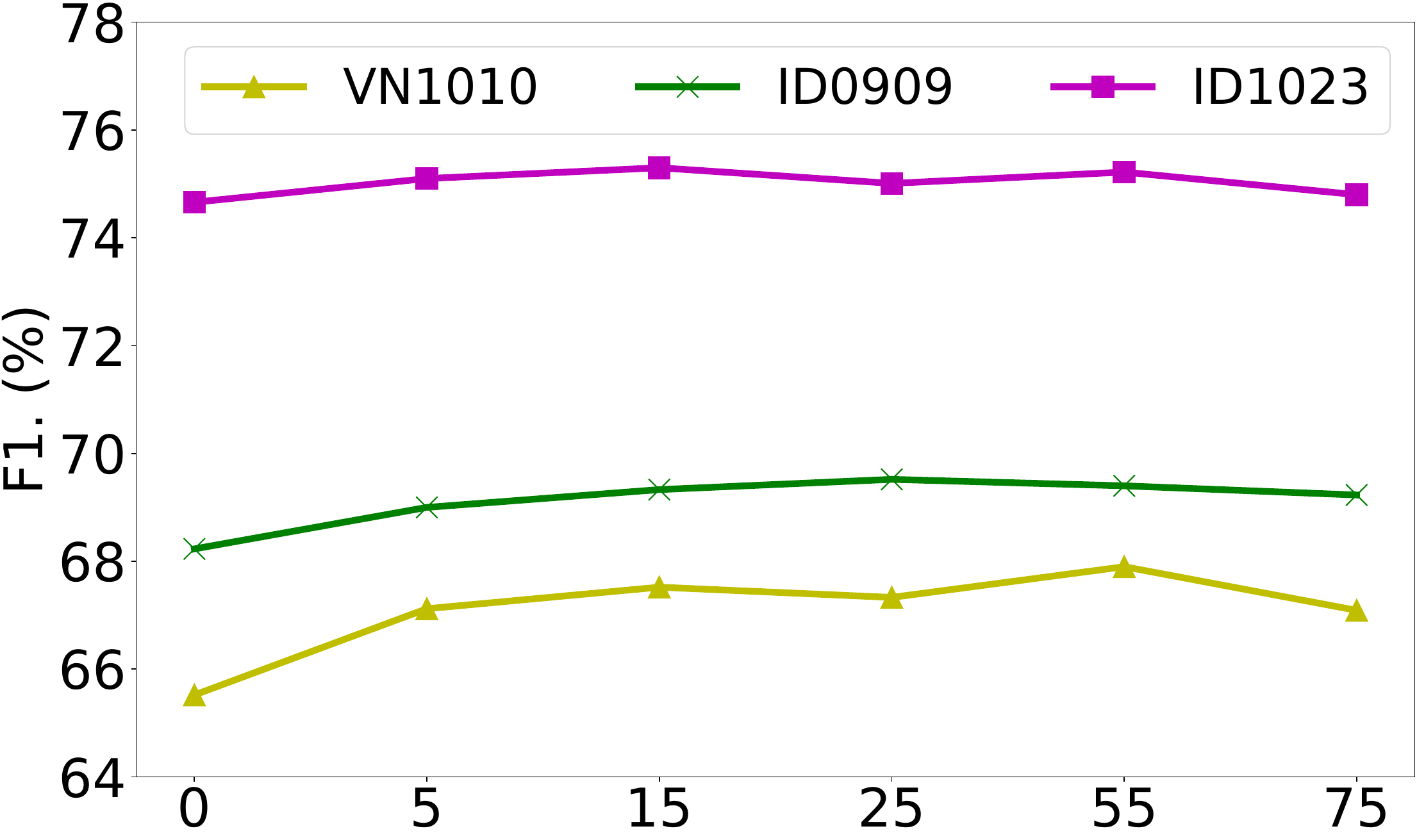}
    \vspace{-5mm}
    \caption{Impact of neighbors for token initialization}
    \label{fig:param_neighbors}
  \end{minipage}
  \hfill
  \begin{minipage}[t]{0.48\linewidth}
    \centering
    \includegraphics[width=1\linewidth]{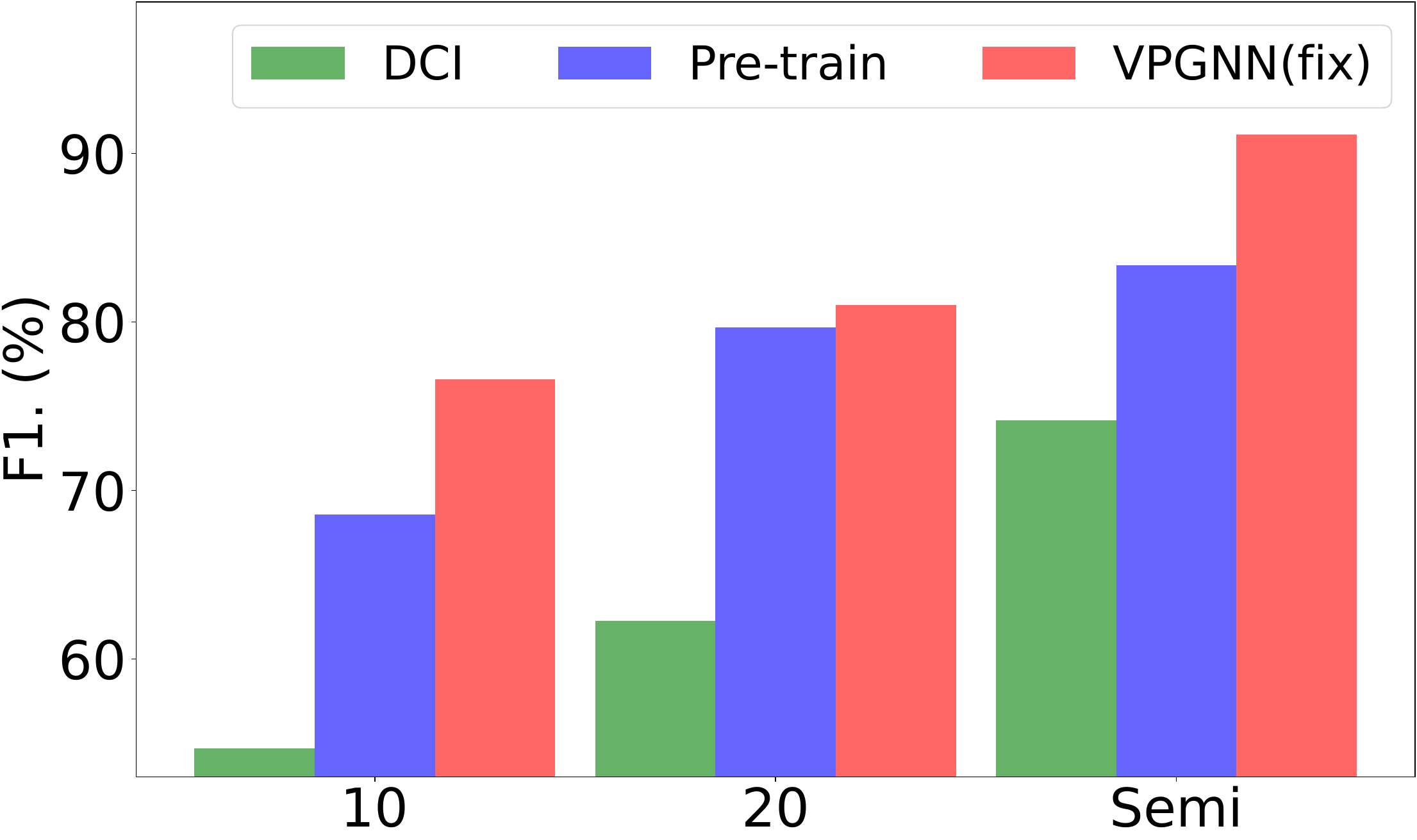}
    \vspace{-5mm}
    \caption{Prompt tuning only without fine-tuning}
    \label{fig:case_f1}
  \end{minipage}
  \label{fig:combined}
\end{figure}

\stitle{Analysis of token initialization.}
When initializing the context tokens, we sample $\eta$ neighbors of the labeled nodes. To analyze the impact of the sampling, we experiment with different $\eta$ values under the 10-shot setting, as shown in Fig.~\ref{fig:param_neighbors}. 
Naturally, when more neighbors are sampled, we observe somewhat better performance. But too many neighbors will also bring more noise. Therefore, it is proper to sample a not big or small amount of neighbors. Besides, we notice that when $\eta=0$, \ie, no neighbor information is used, the performance is worse, reflecting that utilizing neighbor information will make an informative and robust token initialization.

\stitle{Prompt tuning only.}
For large-scale graphs in E-commerce, it is costly to fine-tune the pre-trained GNN model. 
Some studies \cite{li2021prefix, liu2022p} have shown that prompt tuning only while freezing the pre-trained model can still outperform traditional fine-tuning.
To evaluate the effect of only tuning the prompt in \model, we compare it with DCI and Pre-train, two models with fine-tuning on ID1023, across different shots. Specifically, DCI and Pre-train are fine-tuned with 10 epochs, while we fix all pre-trained parameters of \model\ and only tune the prompt vectors. As shown in Fig.~\ref{fig:case_f1}, \model\ still achieves significant improvements over DCI and Pre-train, even when no fine-tuning is done on the pre-trained model. 

\subsection{Online deployment and evaluation}
To further demonstrate the effectiveness of \model\ in production, we deploy it in the Lazada production line to detect voucher abuse orders, and compare it against existing deployed models.

\begin{figure}[t]
\vspace{-2mm}
  \subfigure[Deployment]{
  \centering
  \includegraphics[width=0.475\linewidth]{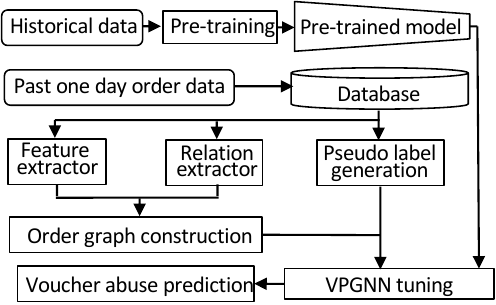}
  }
    \subfigure[Feature engineering]{
  \centering
  \includegraphics[width=0.475\linewidth]{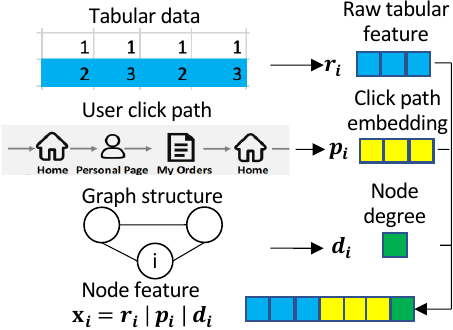}
  }
  \vspace{-4mm}
  \caption{Deployment pipeline and feature engineering.}
  \label{fig:deployment}
\end{figure}

\stitle{Deployment details.}
The deployed model is a \emph{D+1} model, which means that we process and predict the orders placed in the past one day in the production environment. 
Figure~\ref{fig:deployment}(a) describes the deployment pipeline, and (b) zooms into feature engineering.

First, we pre-train a model using historical order data. 
Next, given the past one-day order data flown into the deployed model, they first pass through the feature extractor and relation extractor to exploit rich attributes  and various relations for the order graph construction. Meanwhile, high-confidence pseudo-labels are generated in limited quantity (see Sect.~\ref{sec:method:data}).
Finally, the pre-trained model is loaded and further fine-tuned by \model\ using the pseudo labels.
After tuning, we output the predictions for abusive orders, and a downstream team will handle the predictions appropriately (\eg, canceling orders or suspending accounts).

Moreover, feature engineering still plays a vital role in modern production systems.
First, each order is accompanied by a user click path, in the form of a sequence of actions, \eg, ``\texttt{Home, Personal Page, My orders, Home}'', which is, in fact, a natural language sentence. 
Hence, we choose FastText \cite{bojanowski2017enriching} to model and generate the click path embedding $\vec{p}_{i}$ for order $i$, which is a fast implementation based on the unsupervised skip-gram model, more feasible than large pre-trained language models \cite{devlin2018bert, liu2019roberta} in our high-throughput production setting.
Second, voucher abuse orders tend to have the characteristics of aggregation \cite{liu2018heterogeneous}, \ie,  often have a high degree.
Due to economic constraints, voucher abusers often place a large number of orders from multiple accounts on shared devices and/or IP addresses.
Hence, we also take node degree as an important feature.
Finally, we concatenate the raw tabular feature $\vec{r}_{i}$, the generated FastText embedding of click path $\vec{p}_{i}$ and node degree $d_{i}$ as the node feature $\vec{x}_{i}$ for order $i$.

\stitle{Online performance.}
\model\ is deployed in the \emph{Indonesia} market between \emph{11 and 13 Dec, 2022}, for the Double 12 Campaign. 
We compare \model\ with two existing deployed models: BCP, a distributed implementation of an unsupervised K-means clustering model \cite{sinaga2020unsupervised}, and LPA \cite{raghavan2007near}, an efficient unsupervised community detection algorithm.

As ground truth is only judged on the detected orders, recall cannot be computed since the total number of positive orders is unknown. Besides, the exact precision and the number of true positives cannot be disclosed as they are commercially sensitive. 
Hence, we define a metric called \emph{BCP-normalized Precision Weighted Coverage} (BPWC). Given a model A under evaluation, this metric is calculated by 
\begin{align}
    \small\frac{\text{Precision of model A} \times \text{\# True positives detected by model A}}
    {\text{Precision of BCP} \times \text{\# True positives detected by BCP}}.
\end{align}
Here BCP is treated as the base unit (\ie, 100\%). This metric considers both  precision and true positives (which is proportional to recall), and hence is a good trade-off between more detections and fewer false alerts, without disclosing sensitive information.

As shown in Tab.~\ref{table:online}, compared to BCP and LPA, our proposed \model\ achieves significant improvements in the production setting: over the entire period, \model\ shows a 23.4\% increase over LPA, and almost a 9-fold increase over BCP. The  results demonstrate the advantage of prompt-based fine-tuning on GNNs when dealing with voucher abuse detection.

\begin{table}[t]
   \footnotesize
\addtolength{\tabcolsep}{2pt}
\caption{
Online performance in the Indonesia market over the Double 12 Campaign, 
measured in BPWC.
} 
\label{table:online} 
\begin{tabular}{c|c|c|c|c}  
    \hline
     Model&11 Dec 2022&12 Dec 2022&13 Dec 2022&Overall \\\hline
    BCP &100.0\%&100.0\%&100.0\%&100.0\% \\
    LPA &816.9\%&1784.1\%&330.3\%&809.3\% \\
    \hline
    \model &\textbf{1964.1\%}&\textbf{1990.3\%}&\textbf{469.1\%}&\textbf{998.7\%}\\
    (\% $\uparrow$ over LPA) & (140.4\%) & (11.6\%) & (42.0\%) & (23.4\%)\\
    \hline
\end{tabular}
\end{table}

\section{Conclusion}
In this paper, we proposed a prompt-based fine-tuning approach for GNNs, called \model, to address the problem of voucher abuse detection.
We attempted to bridge the gap between pretext and downstream tasks by proposing a graph prompting function that reformulates the downstream task to follow a similar template as the pretext task. 
The pre-trained GNN model could then be applied with a relatively light fine-tuning given limited downstream labels for abusive orders.  
Extensive offline experiments on both proprietary and public datasets show that \model\ outperforms state-of-the-art baselines in few-shot
and semi-supervised scenarios. Moreover, an online evaluation also demonstrates that \model\ achieves a 23.4\% improvement over two existing deployed models.


\begin{acks}
This research is supported by the Agency for Science, Technology and Research (A*STAR) under its AME Programmatic Funds (Grant No. A20H6b0151).
\end{acks}

\clearpage
\balance
\bibliographystyle{ACM-Reference-Format}
\bibliography{references.bib}

\end{document}